\documentclass[aps,prb,superscriptaddress,twocolumn,showpacs]{revtex4}  
\usepackage{graphicx}
\usepackage{amsfonts}              
\usepackage{amssymb,amsmath}
\usepackage{bm}
\usepackage{comment}
\begin{document}
\title{Unified Semi-Classical Description of Intrinsic Spin-Hall Effect in Spintronic, Optical, and Graphene Systems}
 \author{T. Fujita}       
   \affiliation{Information Storage Materials Laboratory, Electrical and Computer Engineering Department, National University of Singapore, 4 Engineering Drive 3, Singapore 117576}
    \affiliation{Data Storage Institute, A*STAR (Agency for Science, Technology and Research)
DSI Building, 5 Engineering Drive 1, Singapore 117608}
    \author{M. B. A. Jalil}
    \affiliation{Information Storage Materials Laboratory, Electrical and Computer Engineering Department, National University of Singapore, 4 Engineering Drive 3, Singapore 117576}
    \author{S. G. Tan}
    \affiliation{Data Storage Institute, A*STAR (Agency for Science, Technology and Research)
DSI Building, 5 Engineering Drive 1, Singapore 117608}
    \date{\today}    

\begin{abstract}
A semi-classical description of the intrinsic spin-Hall effect (SHE) is presented which is relevant for a wide class of systems. A heuristic model for the SHE is developed, starting with a fully quantum mechanical treatment, from which we construct an intuitive expression for the spin-Hall current and conductivity. Our method makes transparent the physical mechanism which drives the effect, and unifies the SHE across several spintronic and optical systems. Finally, we propose an analogous effect in bilayer graphene.
\end{abstract}
\maketitle
\emph{Introduction.}
The spin-Hall effects\cite{hirsch} (SHE) entail a collection of phenomena in which a pure, transverse spin current is generated in response to a longitudinally applied electric field. Unlike extrinsic versions of the effect,\cite{dyakanov-perel} the intrinsic SHEs are finite in the absence of impurities, originating from the spin-orbit coupling (SOC) present in the band structure of the host system. The rich study of the SHE has been largely fueled by its potential application as a spin current source in the emergent technology called spintronics.\cite{wolf}

The intrinsic SHE has been studied in many systems, especially in semiconductor spintronic systems. Murakami \emph{et al.}\ studied the SHE of holes in $p$-doped bulk semiconductors.\cite{murakami} Independently, Sinova \emph{et al.} studied the effect in two-dimensional electron gases (2DEG) with Rashba SOC.\cite{sinova} Remarkably, the spin-Hall conductivity (SHC) in this system was found to be universal, i.e.\ independent of system parameters, which subsequently stimulated much research of the SHE in 2DEGs (having, in general, both Rashba and linear Dresselhaus SOC).\cite{raimondi,shen} The cubic Dresselhaus SOC case was studied by Bernevig and Zhang.\cite{bern} Two-dimensional hole systems with Rashba SOC also exhibit the intrinsic SHE,\cite{tma,dai,schl} a unique signature being a resonant SHC.\cite{dai} The SHC is typically characterized by the linear Kubo response of the transverse spin current to a charge current excitation. However, this treatment conceals the physical mechanism which drives the effect. The SHE in bulk $p$-semiconductors,\cite{murakami} for example, arises from the Berry curvature of momentum space which results in spin-dependent equations of motion.\cite{sundaram-niu} Similar in nature are the recently proposed SHEs of light,\cite{onoda} phonons,\cite{bliokhph} and excitons.\cite{kuga} On the other hand, the SHE in Rashba 2DEGs occurs as a result of time-resolved spin dynamics induced by an electric field.\footnote{Although distinct, the two mechanism are actually related [\onlinecite{fujita}].} We focus in this paper on the latter type, and hereafter we shall use ``SHE'' to mean this type.

In this paper, we start with a general spin-orbit model and provide a semi-classical picture of the SHE, explaining clearly how it is driven. From there, we construct an intuitive expression for the spin-Hall current and conductivity. Our general expression reproduces exactly the SHC obtained previously using linear response theory in a wide class of systems. Hence our approach not only provides a clear, physical picture for the SHE, but also allows one to classify various SHEs as originating from a common underlying mechanism. Furthermore, our closed expression for the SHC can conveniently be applied to other systems using simple vector algebra, without the need for cumbersome operator algebra.

\begin{figure}[!ht]
\centering
\resizebox{0.5\columnwidth}{!}{
\includegraphics{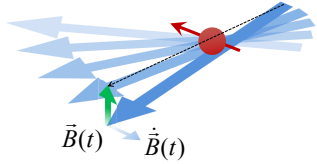}}
\caption{(Color online) In the presence of a time-dependent magnetic field, $\vec{B}(t)=|\vec{B}(t)|\vec{n}(t)$, an additional magnetic field $\vec{B}_\bot=\dot{\vec{n}}\times\vec{n}$ (green, vertical arrow) is seen by spins. The net instantaneous magnetic field felt by spins is the vector sum of $\vec{B}(t)$ and $\vec{B}_\bot$, denoted by the dashed, black arrow.}
\label{stern.fig}
\end{figure}

\emph{Theory.}
%
We consider the general spin-orbit Hamiltonian in the presence of an electric field,
\begin{equation}
\mathcal{H} = \frac{\vec{p}^{\text{ }2}}{2m} - \gamma \vec{\sigma}\cdot\vec{\Omega}(\vec{k})+e\vec{E}\cdot\vec{r},
\label{soc.eq}
\end{equation}
where $\vec{p}=\hbar\vec{k}$ is the momentum, $m$ the effective mass, $\gamma$ is the SOC strength, $\vec{\sigma}$ is the vector of Pauli matrices, $\vec{\Omega}(\vec{k})$ is a momentum-dependent effective field, and $\vec{E}$ is the electric field.  We study the time ($t$) evolution of the above quantum system. To incorporate the explicit $t$-dependence of the system quantum mechanically, we switch to the interaction picture,\cite{townsend} splitting $\mathcal{H}$ into two parts, ${\mathcal{H}} = {\mathcal{H}}_0 + {\mathcal{H}}_1$, where ${\mathcal{H}}_0 = e \vec{E}\cdot\vec{r}$ governs the time evolution of the operators, and ${\mathcal{H}}_1 = \frac{\vec{p}^{\text{ }2}}{2m} -\gamma\vec{\sigma}\cdot\vec{\Omega}(\vec{k})$ governs the time evolution of the states in the new picture. The momentum operator in the interaction picture (subscript $I$) is found to be $\vec{p}_I(t) = e^{i {\mathcal{H}}_0 t/\hbar} \vec{p} e^{-i {\mathcal{H}}_0 t/\hbar} = \vec{p}-e\vec{E}t$, i.e.\ with the expected linear $t$-dependence due to $\vec{E}$. State vectors $|\psi(t)\rangle$ in the Schr\"{o}dinger picture correspondingly transform as $|\psi_I(t)\rangle = e^{i {\mathcal{H}}_0 t/\hbar} |\psi(t)\rangle$, and evolve according to the new ``Schr\"{o}dinger equation'', ${\mathcal{H}}_I(t) |\psi_I(t)\rangle = i\hbar \partial_t |\psi_I(t)\rangle$, where the Hamiltonian ${\mathcal H}_I (t)$ is found to be
\begin{eqnarray}
{\mathcal{H}}_I(t)&=&\frac{\vec{p}_I^{\text{ }2}}{2m}-{\gamma}\vec{\sigma}\cdot \vec{\Omega}(\vec{k}_I(t)),\text{ where}\nonumber\\
\vec{\Omega}(\vec{k}_I(t))&=&\vec{\Omega} - \frac{e E_i t}{\hbar} \frac{\partial \vec{\Omega}}{\partial k_i} + \frac{e^2 E_l E_m t^2}{2 \hbar^2}\frac{\partial^2 \vec{\Omega}}{\partial k_m \partial k_l}-\cdots,\nonumber
\label{interham.eq}
\end{eqnarray}
and the summation over repeated indices is implied. The Hamiltonian ${\mathcal H}_I (t)$ is that of a particle subject to an explicitly $t$-dependent spin-orbit field, $\vec{\Omega}(t)$. We consider diagonalizing the Schr\"{o}dinger equation at time $t$, by applying a unitary rotation $U(t)$ which rotates the $\hat{z}$-axis to be pointing along the instantaneous field $\vec{\Omega}(t)$, i.e.
\begin{eqnarray}
U(t){\mathcal{H}}_I(t) U^\dagger(t)&=& U(t) \left(i\hbar \partial_t \right) U^\dagger(t),\nonumber\\
\frac{\vec{p}_I^{\text{ }2}}{2m}-\gamma \sigma_z |\vec{\Omega}(t)| &=& i\hbar\partial_t+i\hbar U(t) \dot{U}^\dagger(t),\nonumber\\
 &=& {\epsilon}-\hbar\mathcal{A}_0(t).
 \label{hamil_t_eff.eq}
\end{eqnarray}
where ${\epsilon}=i\hbar\partial_t$ is the energy operator. On the right-hand-side, a gauge field $\mathcal{A}_0(t)\equiv -i U \dot{U}^\dagger$ appears from the $t$-dependence of $U$. The term $\hbar\mathcal{A}_0(t)$ represents an additional Zeeman-like term, indicating the presence of an effective magnetic field in the rotating frame. Since $\dot{U}=\frac{1}{2}i U \vec{\sigma}\cdot\vec{\omega}$ [\onlinecite{apoorva}], where $\vec{\omega}$ is the instantaneous angular velocity of the rotating frame, the effective magnetic field in the laboratory frame is found to be $-\vec{\omega}$. Clearly, the angular velocity of the rotating frame is also that of the unit vector $\vec{n}(t)=\vec{\Omega}(t)/|\vec{\Omega}(t)|$, so $\vec{\omega}=\vec{n}\times\dot{\vec{n}}$. Thus, the effective magnetic field arising from the gauge transformation is $\vec{\Omega}_\bot=\dot{\vec{n}}\times\vec{n}$, which is an invariant with respect to the particular choice of $U$, depending only on the $t$-dependence of $\vec{\Omega}(t)$. We illustrate this field in Fig.\ \ref{stern.fig}. A classical derivation of this field can be found in Refs.\ [\onlinecite{fujita,stern}]. The $\vec{\Omega}_\bot$ term represents a physical magnetic field which couples to the electron spins,\cite{stern,xiao} and, as we show below, is precisely the component leading to the SHE.

\emph{Semi-classical calculation of spin-Hall current.}
We consider the DC response of a SOC system to a charge current $j_i$ due to an electric field $E_i$. In particular, we calculate the spin current $j_{s,j}^l = s/2\{v_j,\sigma^l \}$ where $s$ is the value of the spin angular momentum, and also the SHC $\sigma_{ij}^l\equiv j_{s,j}^l/E_i$. The electric field results in a $t$-dependent spin-orbit field $\vec{\Omega}(t)$, which, as we showed above, is always accompanied by an additional field $\vec{\Omega}_\bot=\dot{\vec{n}}\times\vec{n}$ (Fig.\ \ref{stern.fig}). We assume that the spins align themselves to the net magnetic field $\vec{\Omega}_\Sigma$, i.e.\ the sum of $\vec{\Omega}$ and the $t$-dependent correction. The classical spin vector is then $\vec{s} = \pm s\vec{\Omega}_\Sigma /|\vec{\Omega}_\Sigma|$, where $+(-)$ represents spins aligned parallel (anti-parallel) to $\vec{\Omega}_\Sigma$. Along the $\hat{l}$-coordinate, the component of the spin vector due to $\vec{\Omega}_\bot$ is\footnote{The spin may well have a component along $\hat{l}$ due to $\vec{\Omega}$ itself, but it is not proportional to the electric field and does not contribute to the spin-Hall current.} $s^l = \pm s\left(\dot{\vec{n}}\times\vec{n}\right)_l/|\vec{\Omega}_\Sigma|$, where $\vec{\Omega}_\Sigma$ is expressed in terms of its equivalent angular velocity. We now consider the adiabatic limit. In the ideal limit, $|\vec{\Omega}|\rightarrow\infty$, so the spins always remain aligned to it as it varies. In reality, $|\vec{\Omega}|$ is finite and there is a secondary component $\vec{\Omega}_\bot$, and the relevant condition is $|\vec{\Omega}|\gg |\vec{\Omega}_\bot|$, i.e.\ the spins are primarily aligned to $|\vec{\Omega}|$ yet with a small component along $|\vec{\Omega}_\bot|$. The adiabatic spin polarization along $\hat{l}$ is
\begin{equation}
s^l \approx \pm\frac{s\hbar}{2\gamma |\vec{\Omega}|^3}\dot{\Omega}_n \Omega_r \epsilon_{lnr},
\label{sl.eq}
\end{equation}
where $\epsilon_{lnr}$ is the Levi-Civita symbol. We employ the conventional spin current operator, $j_{s,j}^l = s/2\{ v_j,\sigma^l \}$, where $\{ \cdot\}$ denotes the anticommutator. The velocity operator reads  ${v}_j = \partial\mathcal{H}/\partial{p_j}=p_j/m -\gamma\hbar^{-1}\vec{\sigma}\cdot\partial{\vec{\Omega}}/\partial{k_j}$, and upon substituting this expression into the spin current operator we obtain $j_{s,j}^l = s\left(p_j \sigma^l/m-\gamma \partial\Omega_l/\partial{k_j}\right)$. The spin current therefore has two contributions, the first due to the spin-polarization of the states, and second due to the variation in $\vec{k}$ of $\vec{\Omega}(\vec{k})$. The spin-Hall (sH) current is the part of $j_{s,j}^l$ proportional to $E_i$, which corresponds only to the first contribution, thus $j_{s,j}^l (\text{sH}) =  p_j s^l/m$. The total spin-Hall current can be calculated semi-classically by summing the expectation value of the $j_{s,j}^l (\text{sH})$ operator over all states up to the Fermi level,
\begin{equation}
j_{s,j}^l (\text{sH}) = \int \frac{d^D\vec{k}}{(2\pi)^D} \frac{p_j(\vec{k})}{m} s^l(\vec{k}),
\end{equation}
where $D$ is the dimension of the system. We carefully note that the expression for the spin polarization $s^l(\vec{k})$ has opposite signs for spins pointing parallel and antiparallel to $\vec{\Omega}(\vec{k})$ \eqref{sl.eq}. These correspond to the two spin-split bands of the spin-orbit Hamiltonian \eqref{soc.eq}; namely, electrons pointing parallel (antiparallel) to $\vec{\Omega}$ have energies $E_{+(-)}=p^2/2m -(+)\gamma|\vec{\Omega}(\vec{k})|$. Since the Fermi level is usually exceeds the band-splitting, $E_F \gg 2|\vec{\Omega}|$, both bands are occupied with their Fermi wavevectors fulfilling $E_F = \left(\hbar k_F^\pm\right)^2/2m \pm \gamma |\vec{\Omega}(k_F^\pm)|$, where $k_F^- > k_F^+$. Thus, in the region where the two Fermi surfaces overlap (i.e.\ $k < k_F^+$), there is a complete cancellation of $s^l(\vec{k})$. The finite contribution to the spin-Hall current comes from the states in the Fermi surface occupied only by the ground state band, $E_-$. Noting this, and writing the time derivative of $\Omega_n$ in the expression for $s^l$ \eqref{sl.eq} in terms of the wavevector $k_i$, we obtain
\begin{equation}
j_{s,j}^l(\text{sH}) = \int \frac{d^D\vec{k}}{(2\pi)^D} \frac{p_j(\vec{k})}{m} \frac{s\hbar}{2\gamma |\Omega|^3}\left(-\frac{\partial \Omega_n}{\partial k_i}\frac{e E_i}{\hbar}\right)\Omega_r\epsilon_{lnr},
\end{equation}
 where the integration in $k$ goes from limits $k_F^+$ to $k_F^-$. Dividing both sides through by the electric field $E_i$, we obtain an expression for the intrinsic SHC,
\begin{equation}
\sigma_{ij}^l\equiv \frac{j_{s,j}^l(\text{sH})}{E_i}= \frac{es\hbar}{2\gamma m(2\pi)^D}\int d^D \vec{k} \frac{k_j}{|\Omega|^3} \epsilon_{lnr} \Omega_n \frac{\partial\Omega_r}{\partial k_i}.
\label{shc.eq}
\end{equation}
%
\begin{center}
  \begin{table*}
  \label{systems.tab}
  \begin{tabular}{l c c c c}
    \hline\hline 
    System & \hspace{0.2in}$D$\hspace{0.2in} & $\mathcal{H}$ &\hspace{0.2in} $\vec{\Omega}(\vec{k})$ & \hspace{0.2in}$s^z(\vec{k})$ for $\vec{E}=E_x\hat{x}$ \\ \hline \vspace{0.1 in}
    Rashba-Dresselhaus & \hspace{0.2in}$2$\hspace{0.2in} & $\begin{array}{l}\frac{\hbar^2\vec{k}^{2}}{2m}+\alpha\left({k}_x {\sigma}^y - {k}_y {\sigma}^x\right)+\\ \beta\left({k}_y {\sigma}^y - {k}_x {\sigma}^x\right)\end{array}$ &\hspace{0.2in} $\left(\begin{array}{c}\alpha k_y+\beta k_x\\-\alpha k_x-\beta k_y\\0\end{array}\right)$ &\hspace{0.2in} $\mp\frac{e E_x \hbar k_y (\alpha^2-\beta^2)}{4 |\vec\Omega(\vec{k})|^3}$\\ \vspace{0.1 in}
    $k^3$-Dresselhaus & \hspace{0.2in}$3$\hspace{0.2in} & $\begin{array}{l}\frac{\hbar^2\vec{k}^2}{2m}+\eta [ k_x(k_y^2-k_z^2)\sigma^x +\\ k_y(k_z^2-k_x^2)\sigma^y+k_z(k_x^2-k_y^2)\sigma^z ]\end{array}$ &\hspace{0.2in} $\left(\begin{array}{c}-k_x(k_y^2-k_z^2)\\-k_y(k_z^2-k_x^2)\\-k_z(k_x^2-k_y^2)\end{array}\right)$ &\hspace{0.2in} $\mp\frac{e E_x \hbar k_y (k_x^2+k_z^2)(k_y^2-k_z^2)}{4 \eta |\vec{\Omega}(\vec{k})|^3}$\\ \vspace{0.1 in}
    Heavy holes in QW & \hspace{0.2in}$2$\hspace{0.2in} & $\frac{\hbar^2 \vec{k}^2}{2 m} + i \frac{\lambda}{2} (k_-^3 \sigma_+ - k_+^3 \sigma_-)$ &\hspace{0.2in} $\left(\begin{array}{c}k_y^3-3k_x^2 k_y\\k_x^3-3 k_x k_y^2\\0\end{array}\right)$ &\hspace{0.2in} $\mp \frac{9 e E_x \hbar k_y}{4\lambda k^5}$\\ \vspace{0.1 in}
   Bilayer graphene & \hspace{0.2in}$2$\hspace{0.2in} & $-\frac{\hbar^2}{2m}\left( \begin{array}{cc}0 & k_-^2 \\ k_+^2 & 0 \end{array}\right)$ &\hspace{0.2in} $\left(\begin{array}{c}k_x^2-k_y^2\\2 k_x k_y\\0\end{array}\right)$ &\hspace{0.2in} $\tau^z=\mp\frac{2me E_x k_y}{\hbar^2 k^4}$\\
    \hline\hline
  \end{tabular}
    \caption{List of systems and their Hamiltonians $\mathcal{H}$ in which the spin-Hall effect is analyzed. $D$ is the system dimension, $\sigma^l (l=x,y,z)$ are the Pauli spin operators, $k_l$ are the wavevectors, $\sigma_\pm = \sigma^x \pm i \sigma^y$ and $k_\pm = k_x \pm i k_y$. $\vec{\Omega}(\vec{k})$ is the momentum-dependent effective magnetic field, and $s^z$ is the $\tilde{z}$-spin polarization of electrons resulting from an electric field applied in the $\tilde{x}$-direction, obtained using Eq.\ \eqref{sl.eq}. For the case of bilayer graphene, we mean here the pseudospin polarization, $\tau^z$.}
  \end{table*}
\end{center}
%
This is our central result. Using a semi-classical approach, we have constructed a general expression for the intrinsic SHC which makes transparent the physical mechanism driving the phenomena. Our analysis is relevant for describing the SHE in a wide class of systems, as shown below.

\emph{Results and Discussions.}
In Tab.\ \ref{systems.tab} we list several systems which can be represented by the general Hamiltonian in Eq.\ \eqref{soc.eq}. In each case, we give the expression for the $\vec{k}$-dependent magnetic field $\vec{\Omega}(\vec{k})$, and compute the $\hat{z}$-spin polarization, $s^z(\vec{k})$, using Eq.\ \eqref{sl.eq} assuming an electric field applied along $\hat{x}$. Since the $s^z(\vec{k})$ are odd in $k_y$,  carriers traveling in opposite $\hat{y}$-directions become polarized along opposite $\hat{z}$-directions. \emph{This is the physical picture of the SHE.} Below we determine the SHC in each of the spintronic systems. We also examine the optical analog, and finally propose an analogous effect in bilayer graphene.

\emph{A.\ Combined Rashba and Dresselhaus SOC ---}
The out-of-plane spin polarization driven by an electric field in the combined Rashba-Dresselhaus system is given in Tab.\ \ref{systems.tab}. It vanishes when $|\alpha|=|\beta|$, in agreement with previous studies.\cite{shen} When $|\alpha|\neq|\beta|$, using Eq.\ \eqref{shc.eq}, and the relation $k_F^+-k_F^- = -2m|\vec\Omega_{RD}|/\hbar^2$, the SHC is
\begin{equation}
\sigma_{xy}^z =-\frac{e}{8\pi} \frac{\alpha^2-\beta^2}{|\alpha^2-\beta^2|},
\end{equation}
in agreement with Ref.\ [\onlinecite{shen}]. In particular, for the $\alpha=0$ or $\beta=0$ cases, the universal value is reproduced.
%

\emph{B.\ $n$-doped bulk semiconductors ---}
For conduction electrons under the influence of $k^3$-Dresselhaus SOC, the SHC \eqref{shc.eq} is
\begin{eqnarray}
\sigma_{xy}^z &=& \frac{e\hbar^2}{4 \eta m(2\pi)^3}\int_0^{2\pi} \int_0^\pi \int_{k_F^+}^{k_F^-} dk d\theta d\phi\times\nonumber\\
 &\times &\left( k^2 \sin\phi \frac{k_y^2\left( k_z^2-k_y^2\right)\left( k_x^2+k_z^2\right)}{|\vec{\Omega}_D|^3}\right),
\end{eqnarray}
where $\vec{\Omega}_D$ is the $k^3$-Dresselhaus field, and $(\theta,\phi)$ are spherical angles in $\vec{k}$-space. Using the interband relation $k_F^+ - k_F^- \approx -2m|\vec{\Omega}(k_F)|/\hbar^2 k_F$, where $k_F = (k_F^+ + k_F^-)/2$, we produce the SHC of $\sigma_{xy}^z=k_F/12\pi^2$ as computed in Ref.\ [\onlinecite{bern}].

\emph{C.\ Holes in III-V semiconductor quantum wells with Rashba SOC ---}
The SHC \eqref{shc.eq} is $(s=3/2)$, 
\begin{eqnarray}
\sigma_{xy}^z &=& -\frac{9 e \hbar^2}{4\lambda m(2\pi)^2} \int_0^{2\pi} \int_{k_F^+}^{k_F^-} dk d\phi \frac{\sin^2\phi}{k^2},\nonumber\\
&=& -\frac{9 e \hbar^2}{16\pi \lambda m}\left( \frac{1}{k_F^+} - \frac{1}{k_F^-} \right),
\end{eqnarray}
where $\phi$ is the azimuthal angle in $\vec{k}$-space. In the limit of small Rashba coupling, the band resolved Fermi wavevectors can be shown to be $k_F^\pm \approx\sqrt{2\pi n}\mp (2\lambda m/\hbar^2)\pi n$, where $n$ is the hole density, giving a universal SHC of $\sigma_{xy}^z=-9e/8\pi$.\cite{schl} In comparison to the value of $-e/8\pi$ in 2DEGs, the extra factor of $9$ here arises from the larger angular momentum of heavy holes, and the $k^3$-dependence of the spin-orbit coupling.

\emph{D.\ Rayleigh scattering of polaritons ---}
An analogous effect occurs in optics, when polaritons undergo Rayleigh scattering.\cite{kavokin} The polariton polarization  is represented by a pseudospin $\vec{\tau}$, where the pseudospin field is exactly that of bilayer graphene (see below). Upon scattering (which changes the wavevector $\vec{k}$), polaritons acquire a finite $\tau^z$ component, corresponding to circular polarization, whose sign depends on the initial $\vec{k}$. This is the optical SHE.\cite{kavokin}

\emph{E.\ Bilayer graphene ---}
Finally, we propose an analogous effect in bilayer graphene (BG). The BG system is modeled as two coupled honeycomb lattices, with each layer having two inequivalent lattice sites $\tilde{A}, \tilde{B}$ and $A, B$ in the top and bottom layers respectively. We assume the Bernal stacking ($\tilde{A}$-$B$) configuration. In the low energy limit, electrons in the BG system are described by an effective $2$-by-$2$ Hamiltonian,\cite{mccann} $\mathcal{H}_{BG} =-\frac{\hbar^2}{2m}\vec{\tau}\cdot\vec{\Omega}_{BG}$, where $\vec{\Omega}_{BG}=(k_x^2-k_y^2,2 k_x k_y,0)$. The $\vec{\tau}$ here is the vector of Pauli operators acting on the pseudospin, rather than the actual electron spin. The BG eigenstates $\vec{\Psi}$ are two component wavefunctions describing the electronic amplitude of electrons on the two layers, $\vec{\Psi} = (\psi(A),\psi(\tilde{B}) )$. When an electric field $\vec{E}=E_x \hat{x}$ is applied to the bilayer system, an additional out-of-plane component accompanies the strictly in-plane pseudospin field $\vec{\Omega}_{BG}$. The out-of-plane component induces in the adiabatic limit a pseudospin polarization along $\hat{z}$ (see Tab.\ \ref{systems.tab}), which corresponds to charge transfer between the two layers. This is an essential ingredient for the technology known as \emph{pseudospintronics},\cite{min} in which binary states are encoded by relative charge densities on the two BG layers.
\begin{figure}[!ht]
\centering
\resizebox{0.8\columnwidth}{!}{
\includegraphics{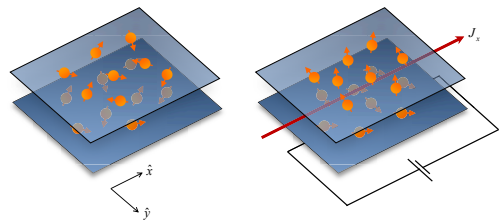}}
\caption{(Color online) Illustration of proposed pseudospin-Hall effect in bilayer graphene. The arrows indicate the direction of the electron momenta. (left) With no electric field, electrons with all momenta are distributed evenly between the two layers. (right) With an applied electric field in the $\hat{x}$-direction, electrons are separated to each layer depending on their $\hat{y}$-momentum; electrons with $+(-) p_y > 0$ are transferred to the bottom (top) layers respectively.}
\label{pshe.fig}
\end{figure}
The physical effect of this polarization is illustrated in Fig.\ \ref{pshe.fig}: electrons with $p_y > 0$ ($p_y < 0$) are separated to the bottom (top)  graphene monolayer of the BG system. The effect is completely analogous to the SHE with replacements $\vec{\sigma}\leftrightarrow\vec{\tau}$. It may therefore be called the \emph{pseudospin-Hall effect}. Such an effect should be of interest to the field of pseudospintronics.

We finally reiterate that there exists a second class of the intrinsic SHE arising from the non-trivial curvature of $\vec{k}$-space, which was not discussed in this paper.\cite{murakami,onoda,bliokhph,kuga} We have shown elsewhere that the underlying mechanisms, although distinct, are related.\cite{fujita}

\end{document}